# Airy-like hyperbolic shear polariton in high symmetry van der Waals crystals

*Yihua Bai,[1,†] Qing Zhang,[1,†,*] Tan Zhang,[2] Haoran Lv,[1] Jiadian Yan,[1] Jiandong Wang,[1] Shenhe Fu,[3] Guangwei Hu,[4] Cheng-Wei Qiu[2,*] and Yuanjie Yang,[1,*]*

[1] School of physics, University of Electronic Science and Technology of China, Chengdu 611731, China

[2] Department of Electrical and Computer Engineering, National University of Singapore, Singapore 117583, Singapore

[3] Department of Optoelectronic Engineering, Jinan University, Guangzhou 510632, China

[4] School of Electrical & Electronic Engineering, Nanyang Technological University, 639798, Singapore

†These authors contributed equally to this work.

*E-mail: qingzhang@uestc.edu.cn; chengwei.qiu@nus.edu.sg; dr.yang2003@uestc.edu.cn

**Abstract**

Controlling light at the nanoscale by exploiting ultra-confined polaritons – hybrid light and matter waves – in various van der Waals (vdW) materials empowers unique opportunities for many nanophotonic on-chip technologies. So far, mainstream approaches have relied interfacial techniques (e.g., refractive optics, meta-optics and moire engineering) to manipulate polariton wavefront. Here, we propose that orbital angular momentum (OAM) of incident light could offer a new degree of freedom to structure vdW polaritons. With vortex excitations, we observed a new class of accelerating polariton waves – Airy-like hyperbolic phonon polaritons (PhPs) in high-symmetry orthorhombic vdW crystal α-MoO$_3$. In analogous to the well-known Airy beams in free space, such Airy-like PhPs also exhibit self-accelerating, nonspreading and self-healing characteristics. Interestingly, the helical phase gradient of vortex beam leads to asymmetry excitation of polaritons, as a result, the Airy-like PhPs possess asymmetric propagation feature even with a symmetric mode, analogous to the asymmetry hyperbolic shear polaritons in low-symmetry crystals. Our finding highlights the potential of OAM to manipulate polaritons in vdW materials, which could be further extended into a variety of applications such as active structured polaritonic devices.

Polaritons, quasi-particle half-light and half-matter excitations, in van der Waals (vdW) materials that exhibit long lifetimes, low loss and strong field confinement[1], offer a unique atomically thin platform for future compact and integrated photonics devices. A vibrant research field has been thereby emerging, broadly dubbed as interface polaritonics[2], to extremely manipulate vdW polaritons for advanced nanophotonic and opto-electronic applications[3-7]. Intriguing phenomena have been demonstrated, such as refractive optics for polariton steering and focusing[8-10], meta-optics for wavefront control[11, 12], moiré engineering for topological polaritons[13-16], and photonic crystal nano-light[17]. Yet, so far many of these pioneering studies are from the material strategies or geometric strategies to control vdW polaritons, the wavefront lacks tunability once the structure is fabricated. In these systems, polaritons are excited by regular dipole source or plane wave, there has been limited channel to manipulate the vdW polariton wavefront from incident properties. Hence that, exploring additional degrees of freedom to actively control such polariton waves are urgently demanded.

It's well known that light is able to carry orbital angular momentum (OAM), which are quantized by $L = l\hbar$ per photon ($l$ is the topological charge, $\hbar$ is the reduced Planck constant)[18-20]. Exploiting the OAM modes of light has been widely developed to applications, ranging from high-capacity optical communication[21], optical tweezer[22, 23], to quantum entanglement[24]. However, most OAM-induced phenomena are limited to isotropic materials, and directly using OAM for tunable vdW polaritonic devices has been rarely reported. In this regard, the interplay between OAM of light and anisotropic polaritons is largely unknown and diverse, and remain to be revealed in a clear, and unambiguous manner.

In particular, in-plane hyperbolic phonon polaritons (PhPs) have been found in orthorhombic crystals, such as α-MoO$_3$ and α-V$_2$O$_5$[25,26]. Compared to metal plasmons that suffer from high optical loss and valley excitons that feature short lifetime at room temperature, the PhPs can propagate with ultralong lifetime and ultrahigh field confinement, which advance for on-chip long-range transport of optical information[27, 28]. Here, we show that the interplay between vortex beams and extreme anisotropy of vdW materials yields the generation of Airy-like hyperbolic PhPs in high-symmetry orthorhombic crystal. The Airy beam shifts along a curved self-accelerating trajectory[29] which exhibits desirable attributes such as diffraction-free[30, 31] and self-healing[32]. These exotic properties thus have been utilized for applications in the areas of biomedical imaging[33] and particle manipulation[34], etc. To date, accelerating waves have been proposed in water[35], free-space optics[36], and in the form of SPPs[37, 38] and electrons[39], whereas no such Airy wave has been reported in vdW polariton systems.

In this work, we theoretically and numerically demonstrate the emergence of Airy-like PhPs in high symmetry crystal α-MoO$_3$. We use a gold disk to efficiently launch PhPs via vortex excitation. By finite-difference time-domain (FDTD) simulations, we found nontrivial interference pattern of PhPs waves, where many lobes of parabolic trajectory have been generated at two side of the disk. After theoretically analysis the phase modulation of OAM, along with the hyperbolic propagating nature of PhP waves, we clear that such parabolic lobes are exactly Airy-like PhPs, where a cubic phase modulation (the nature of 2D Airy wave packet[29, 30, 40]) is satisfied. The parabolic trajectory of PhP waves can be largely tailored by the topological charge of the OAM beams, where the deviation of main lobe extends to several micrometers. Moreover, owing to the helical phase gradient of OAM, the Airy-like PhPs possess asymmetric propagation in α-MoO$_3$, analogous to the asymmetry PhPs in low-symmetry crystals. This is a more flexible approach that only utilizing vortex beams to realize hyperbolic shear polariton even in high symmetry crystals, while PhPs in low-symmetry (e.g., monoclinic and triclinic) crystals are largely restricted by their naturally occurring permittivities. Our work provides an effective method to realize structured polaritons (e.g., Airy-like PhPs) and hyperbolic shear polaritons, motivating new opportunities for on-chip OAM-based polaritonic devices.

The schematic of our device to generate Airy-like PhPs is shown in Fig. 1a. Biaxial orthorhombic crystal α-MoO$_3$ features in-plane anisotropy (e.g., $\varepsilon_{100}$ < 0 and $\varepsilon_{001}$ > 0) at its Reststrahlen bands (RBs), leading to in-plane hyperbolic isofrequency curve (IFC) of supported PhPs in momentum space (Fig.1c). Due to the in-plane hyperbolicity, a metallic micro-disk can be used as planar hyperlens to focus the PhPs at outside of the disk (Fig. 1b) [41-43]. Because of the high-symmetry of orthorhombic crystals, the PhPs propagate symmetrically away from the disk with mirror-symmetric and focused at the central horizontal line. However, these hyperlens are excited by Gaussian plane waves. In principle, it is feasible to bring an extra phase to PhP waves by introducing a phase variation to the incidence directly, as we proposed in Fig. 1a. Hence that, the PhPs interfere with each other and ultimately form the Airy-like PhPs at the outside of the disk (Fig. 1d), demonstrating a self-bending wave with multiple lobes. In addition, this Airy-like PhPs are one kind of hyperbolic shear polaritons, as revealed by the Fourier transform of the asymmetric electric field (Fig. 1e), exhibiting a stronger intensity along one side of the hyperbola.

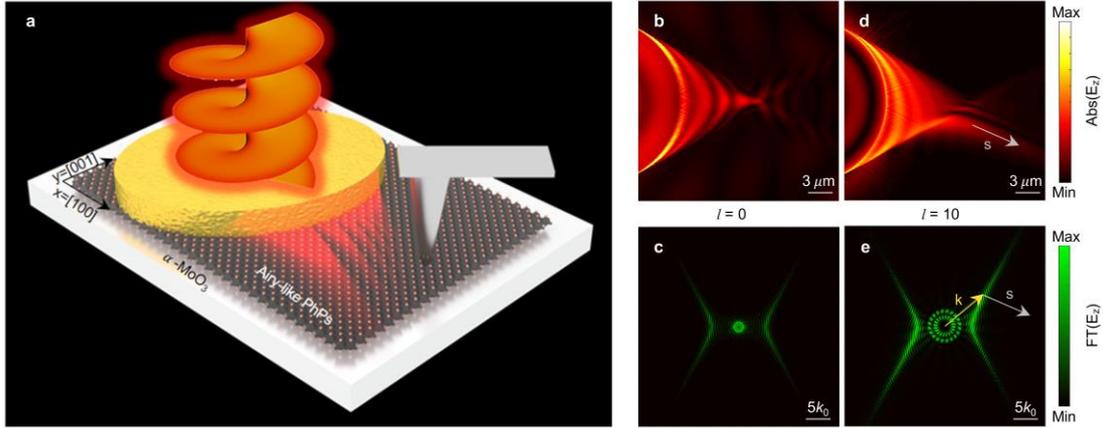

**Figure 1. Schematic of the Airy-like PhPs.** (a) The PhPs in high symmetry α-MoO$_3$ flake excited by vortex excitation. (b, d) Simulated electric fields [abs(E$_z$)] of PhPs excited without and with OAM ($l$=10) at frequency 910 cm$^{-1}$. (c, e) Fourier transformation (FT) of the electric field in (b, d). The radius of gold disk is set as 10 μm. The thickness of α-MoO$_3$ slab is 150 nm, x and y correspond to the [100] and [001] crystalline directions of α-MoO$_3$, respectively.

To clear the generating principle of such nontrivial Airy-like PhPs, we propose an interference model based on Huygen's principle in hyperbolic vdW materials as illustrated in Fig. 2a, where the red dashed and solid lines represent the wavefronts of excited PhPs. The gold disk can be regarded as an antenna consists of an infinite number of dipole sources with continuous phase difference located along its circular periphery when the vortex incidence is employed. The launched PhPs carry extra azimuthal phase gradient, thereby driving the focal spots deviate from the [100] optical axes of α-MoO$_3$ (more details, see supplementary Section S2, Fig. S2). Figure 2b shows the PhPs excited by a vortex incident beam ($l$ = 25) at frequency 910 cm$^{-1}$, where the interference trajectories are in good agreement with the theoretical result in Fig. 2a

(green solid lines).

In particularly, the focal length of such a hyperlens can be analytically calculated as $f = R\sqrt{1 - \varepsilon'_x/\varepsilon'_y} \approx 20$ μm in this model by illuminating plane waves [43], where $R$ is the radius of the gold disk, $\varepsilon'_x$ and $\varepsilon'_y$ are real parts of permittivity of the α-MoO$_3$ material for *x* and *y* components, respectively. Generally, a 3-power phase pattern can be required to generate Airy beams [29]. As shown in Fig. 2c, the phase along the vertical line at the focus (white dashed line in Fig. 2b) is extracted. It is obviously seen that the simulated data from the electric field matches well with the 3-power phase fitting curve, manifesting the generation of Airy-like PhPs (for more details, see supplementary Section S2, Fig. S3).

We further analyze the spatial evolution of aforementioned Airy-like PhP waves. Supposing the α-MoO$_3$ surface is located at z = 0 and x is the propagating direction, the electric field of the in-plane PhPs can be expressed by the Helmholtz equation

$$\nabla^2 E + k_0^2 E = 0 \tag{1}$$

The electric field can be written as $E(x, y, z) = A(x, y)\exp(ik_x x)\exp(-\gamma z)$, where $\gamma^2 = k_x^2 - k_0^2$, $k_x$ is the wavenumber of the PhPs along the x direction (More details about dispersion relation of the biaxial α-MoO$_3$, see Supplementary Information Section 1). The wavefunction of PhPs can be further deduced to the form of Schrödinger equation

$$\frac{\partial^2 A}{\partial y^2} + 2ik_x \frac{\partial A}{\partial x} = 0 \tag{2}$$

In order to investigate the trajectory of PhPs, we consider the initial wavefunction $A(x = f, y) = \text{Ai}[y/y_0]\exp(ay/y_0)\exp(ivy/y_0)$, where $\text{Ai}[y/y_0]$ denotes the Airy function, $y_0$ is the characteristic length scale, $a$ is the function defining the exponential apodization and $v$ is related to the initial launch angle $\theta_0$ by $\theta_0 = v/(k_x y_0)$ [44]. The solution for the PhP wave can be written as

$$A(x, y) = \text{Ai}\left[\frac{y}{y_0} - \left(\frac{x-f}{2k_x y_0^2}\right)^2 - \frac{v(x-f)}{k_x y_0^2} + ia\frac{x-f}{k_x y_0^2}\right] \exp\left[a\frac{y}{y_0} - \frac{a}{2}\left(\frac{x-f}{k_x y_0^2}\right)^2 - \frac{av(x-f)}{k_x y_0^2}\right] \tag{3}$$

$$\times \exp\left[i\left(-\frac{1}{12}\left(\frac{x-f}{k_x y_0^2}\right)^3 + \left(a^2 - v^2 + \frac{y}{y_0}\right)\frac{x-f}{2k_x y_0^2} + \frac{vy}{y_0} - \frac{v}{2}\left(\frac{x-f}{k_x y_0^2}\right)^2\right)\right]$$

It is clear that PhP wave can follow a trajectory in the x-y plane described by a parabola $y = y_0((x-f)^2/4k_x^2 y_0^4 + v(x-f)/k_x y_0^2)$. An ideal diffraction-free Airy beam possesses infinite energy, and PhP Airy waves with exponential apodization profiles could still exhibit their remarkable characteristics.

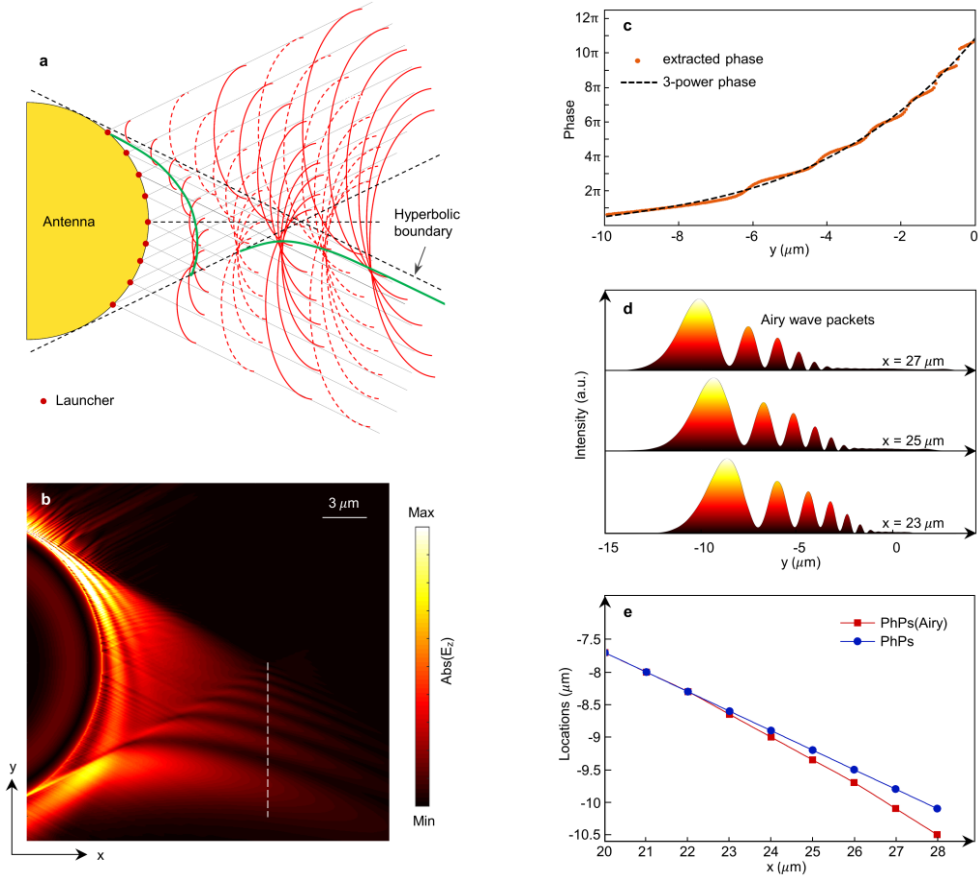

**Figure 2. The Airy wave packet feature of OAM excited PhPs.** (a) The schematic based on Huygen's interference model. (b) Simulated abs($E_z$) of PhPs, showing the self-accelerating trajectory in the x-y plane. (c) Phase distribution along the vertical line at focus (the white dashed line in (b)) together with the 3-power phase modulation. (d) Airy wave packets at x = 23 μm, 25 μm, 27 μm, respectively. The Airy-like PhPs remain almost non-spreading. (e) The self-accelerating trajectory of the Airy wave. The blue line is the linear trajectory of PhPs through dipole excitation.

Analogous to Airy beams in free space, as shown in Fig. 2d, such PhP Airy wave packets remain almost non-spreading during propagation at x=23, 25, 27 μm, respectively. Besides, there exists a larger shift between the main lobes with a larger x, which can be understood as varying group velocities of Airy wave packets along propagation, known as self-accelerating property. To further confirm its Airy wave feature, locations of the main lobes of the Airy wave (red curve) and wavefronts of PhP wave excited by dipole (blue curve) are compared, as demonstrated in Fig. 2e. Distinct parabolic trajectory of Airy wave is illustrated, whereas PhP wave by dipole excitation presents a linear trajectory as a comparison.

Another prominent propagation attribute of Airy beams is the self-healing property, namely, the Airy beams can be reconstructed when the main lobe is initially perturbed. To better validate the interference model in Fig. 2a, dipole array with phase delay is used as the incident source. Using the same parameters above, Figure 3(a, c) show simulated results of PhPs excited by z-polarized dipole array ($\Delta \phi = 15 * 2\pi$), which can be regarded as an equivalence to the vortex beam (*l*=15). Fig. 3(b, d) present the results with a circular obstruction (the diameter is 1μm) placed on the α-MoO3 surface in front of the main lobe. It is evident that the parabolic trajectory reconstructs by itself after the perturbation, revealing typical self-healing property of Airy beams (more cases with varying-sized obstacles, see section S7 Fig. S8).

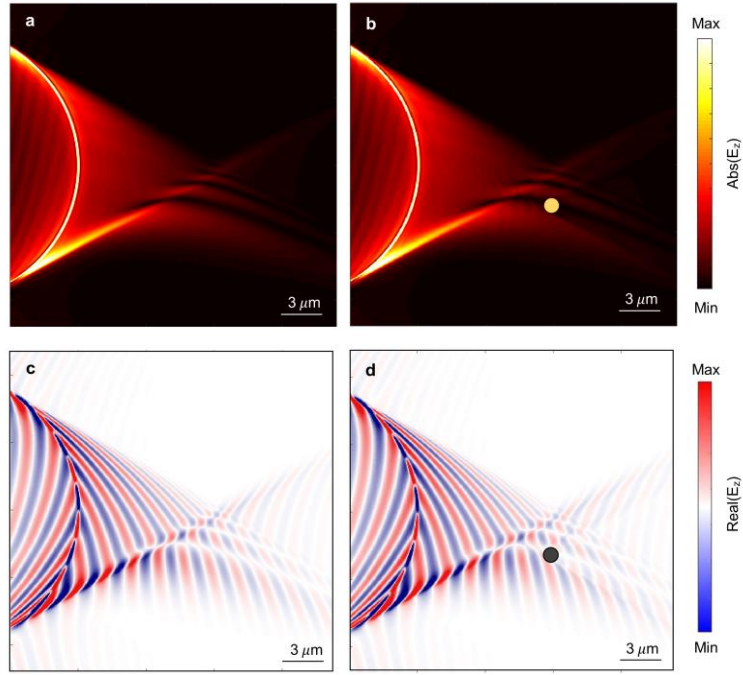

**Figure 3. Self-healing property of Airy-like PhPs.** Simulated results of launched PhPs without (a, c) and with (b, d) an artificially placed block on the α-MoO3 surface in front of the main lobe. (a, b) The simulated electric field distributions excited by z-polarized dipole array with phase delay (*l*=15). (c, d) The corresponding wavefronts of PhP waves.

In order to demonstrate the tunability of Airy-like PhPs, numerical simulations with different OAM states are performed. Figure 4 shows stimulated PhPs by illuminating radially polarized Laguerre–Gaussian beams carrying different topological charges (*l* = -10, 0, 10, 20) at 910 cm$^{-1}$. For traditional Gaussian beam (*l* = 0), the propagating PhPs interfere and form a focal spot (Fig. 4b). As shown in Fig. 4f, the IFC is symmetric and the angle between the asymptotes of the hyperbola can be described as $\alpha = 2\arctan\sqrt{-\varepsilon_x/\varepsilon_y}$. In Fig. 4c, with vortex excitation, the electric field begins to incline towards the lower side and reveals a self-accelerating trajectory. For a vortex beam with opposite topological charge (Fig. 4a), the phase difference is inverse and the electric field inclines to the upper side accordingly. Figure 4g and 4e show that the corresponding IFCs become asymmetric, where the yellow and grey arrows are wavevector and Poynting vector, respectively. The Poynting vector determine the energy flow of excited PhPs, which is perpendicular to the wave vector in such hyperbolic materials.

Interestingly, the Airy-like PhP manifest itself as one kind of hyperbolic shear polaritons. One crucial feature is the asymmetric intensity along two sides of the hyperbola in Fig. 4(e-h). The asymmetric shear polaritons emerge on the hyperbolic material with an incident vortex beam serving as the excitation source, further confirming the hypothesis to generate asymmetric hyperbolic shear polaritons in high-symmetry orthorhombic crystal without off-diagonal elements[45]. As illustrated in Fig. 4h, the asymmetry of IFCs becomes more remarkable with an increasing topological charge.

As the focal length is irrelative to the OAM carried by incident beams, we plot the electric field along the vertical line at the focal spot to further characterize the Airy-like response. As shown in Fig. 4i and 4j, the deviation of the main lobe (black arrows) is dependent on the topological charge. With the absolute value of OAM order changing from 10, 15, 20 to 25, the main lobes deviate 2.9 μm, 4.6 μm, 5.8 μm and 7.8 μm, respectively (Fig. 4k). On the basis of this method, the OAM state, in turn, can be accurately deduced by measuring the deviation of the main lobe. This approach enables a dynamically tunable manipulation of PhPs by varying the topological charge of incident light, exemplifying a generalizable framework to manipulate the flow of nanolight in many other polaritonic systems.

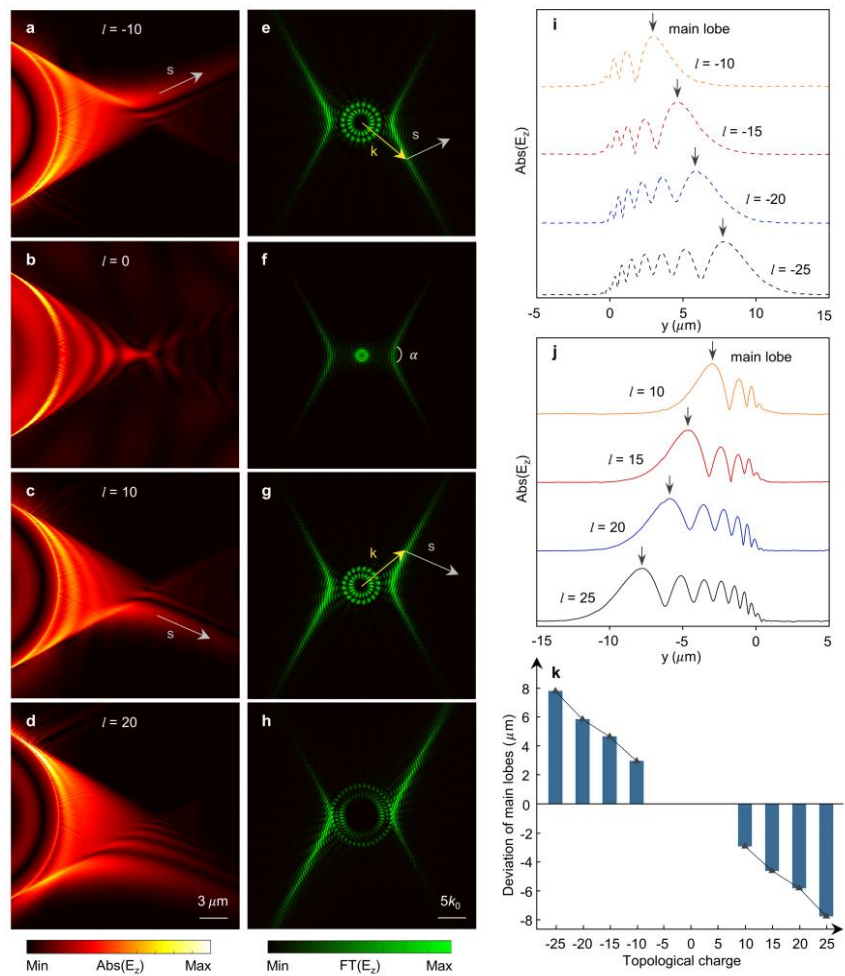

**Figure 4. Tunable excitation of Airy-like hyperbolic shear polaritons based on the OAM.** (a-d) Field distributions of PhPs with different topological charges ($l$= -10, 0, 10, 20). (e-h) The corresponding Fourier transformation of electric fields in (a-d). And simulated amplitude distributions along the vertical line at the focal spot, with OAM order ranging from (i) -25 to -10, (j) 10 to 25. (k) Deviation of the main lobes with respect to various OAM orders. The solid black triangles represent the shifting distances.

In conclusion, we have demonstrated a novel accelerating polariton waves – the Airy-like PhPs in high-symmetry orthorhombic vdW material α-MoO$_3$ excited by vortex beam. The generated PhPs reveal Airy features (e.g., non-diffracting, self-bending and self-healing characteristics) within their propagation. Notably, the topological charge of incidence can be utilized to actively control the parabolic trajectory of the Airy-like PhP waves. Meanwhile, such Airy-like PhPs possess hyperbolic shear polariton feature even in a high symmetric crystal. Our work opens up new avenues for controlling light at the nanoscale, and hold potential applications for many nanophotonic technologies such as actively polaritonic on-chip devices.

**Acknowledgments**

This work was supported in part by the National Natural Science Foundation of China (Nos. 12174047, 62205049 and 11874102), the Sichuan Province Science and Technology Support Program (No. 2020JDRC0006), and by the Sichuan Science and Technology Program (Nos. 2022YFH0082 and 2022YFSY0023). Q.Z. acknowledges the support from the start-up funding of University of Electronic Science and Technology of China.